\documentclass
[preprint,showpacs,byrevtex,10pt,twocolumn,tightenlines,prl,letterpaper]{revtex4}%
\usepackage{amsfonts}
\usepackage{amsmath}
\usepackage{amssymb}
\usepackage[dvips]{graphicx}
\usepackage{hyphenat}%
\providecommand{\U}[1]{\protect\rule{.1in}{.1in}}
\begin{document}
\preprint{ }
\title{Microstructure and the Boson-peak in thermally-treated In$_{\text{x}}$O films}
\author{Itai Zbeda$^{1}$ and Ilana Bar$^{1}$ and Z. Ovadyahu$^{2}$}
\affiliation{$^{1}$Physics Department, Ben-Gurion University of the Negev, Beer-Sheva,
8410501 Israel}
\affiliation{$^{2}$Racah Institute of Physics, The Hebrew University, Jerusalem 9190401,
Israel }

\begin{abstract}
We report on the correlation between the boson-peak and structural changes
associated with thermally-treating amorphous indium-oxide films. In this
process, the resistance of a given sample may decrease by a considerable
margin while its amorphous structure is preserved. In the present study, we
focus on the changes that result from the heat-treatment by employing
electron-microscopy, X-ray, and Raman spectroscopy. These techniques were used
on films with different stoichiometry and thus different
carrier-concentration. The main effect of heat-treatment is material
densification, which presumably results from elimination of micro-voids. The
densified system presents better wavefunction-overlap and more efficient
connectivity for the current flow. X-ray, and electron-beam diffraction
experiments indicate that the heat-treated samples show significantly less
spatial heterogeneity with only a moderate change of the radial-distribution
function metrics. These results are consistent with the changes that occur in
the boson-peak characteristics due to annealing as observed in their Raman spectra.

\end{abstract}
\maketitle

\section{Introduction}

Disorder plays a major role in the properties of solids. The study of disorder
is a challenge in terms of being able to control, characterize, and quantify
it. A significant effort in this vein was invested in the field of electronic
transport. In particular, the need for modifying and quantifying disorder is
an essential ingredient in the field of disorder-induced phenomena. Prime
examples in this category are the metal-insulator transition and the
superconductor-insulator transition. The system resistivity is sometimes used
as an empirical measure of disorder in these studies. Electric conductivity of
a solid is arguably its most sensitive property and it may be affected by
different means, not all of them may be attributed to disorder. A change in
carrier-concentration for example, naturally affects conductivity while only a
small (and indirect) change in disorder may be incurred in the process.

Restricting the lateral dimensions of the sample has been widely used to
increase scattering. This is in particular a viable technique to change
disorder in transport studies of thin films. However to introduce strong
disorder in a three-dimensional system one may have to resort to alloying or
use a two-component mixture; a granular system. An effective method that was
employed to introduce disorder in a metallic system was exposure to neutron or
$\alpha$-particle radiation. This technique was employed in modifying the
transport properties of A-15, MgB$_{\text{2}}$, and other materials by
introducing point-defects and grain-boundary spacing \cite{1}.

A complementary, backward procedure is thermal-annealing. This is an effective
way to reduce disorder when the system is disordered to start with, a common
situation in vacuum deposited or quench-cooled systems. Thermal-annealing has
been successfully used in various transport experiments as a means of
fine-tuning the disorder of amorphous indium-oxide (In$_{\text{x}}$O) films.
It has been shown that thermally-treating In$_{\text{x}}$O films may result in
resistance change of up to 4-5 orders of magnitude at room-temperature with
only a small change of carrier-concentration measured by the Hall effect. It
seems therefore that the huge change in resistance is due to enhanced
mobility, suggestive of a less disordered system. The range of disorder
attainable with this method allows studies of both sides of the
metal-insulator, and the superconductor-insulator transitions of this
material. It was argued that the main reason for the resistance change is
densification \cite{2}. This was supported by demonstrating the similarity of
the change in the optical-gap during thermally-annealing In$_{\text{x}}$O
films, and in studies of pressure-induced densification of glasses.

Another property known to be sensitive to disorder in solids is the boson-peak
(BP), which has been widely studied in amorphous systems and glasses
\cite{3,4,5,6,7,8,9,10,11,12}. The BP is a feature that appears at the low
energy vibrational-density-of-states of amorphous and disordered systems. This
feature may be resolved by heat-capacity, neutron scattering experiments, and
by Raman spectroscopy. There are several competing theoretical models that
purport to account for the mechanism that underlies the BP. The common
ingredient in the great majority of them is spatial disorder
\cite{4,5,8,13,14,15,16}. This makes the characteristics of the BP a relevant
probe for monitoring changes in disorder. This is of particular relevance for
amorphous systems where quantifying disorder is a challenge. In addition to
the lack of long-range order, most amorphous solids exhibit mass density that
is lower than their crystalline counterpart. This is presumably due to their
being formed by fast cooling from the liquid or gaseous phase \cite{17}. The
latter preparation method, quench-cooling the material from the vapor phase
onto a cold substrate, usually results in a porous structure that has many
micro-voids reducing the material bulk specific gravity. Consequently, an
appreciable volume change may be affected in these structures upon application
of pressure, and by thermal-treatment \cite{2}.

In this work we attempt to further elucidate these issues using several
versions of amorphous indium-oxide as a model system for a metallic glass.
This allows us to track changes in the BP due to quantifiable changes of
disorder. Raman spectroscopy taken from as-made and annealed In$_{\text{x}}$O
films of different composition, reveal significant changes in the BP magnitude
and shape. It is noted that Raman spectroscopy may not faithfully convey the
detailed shape of the BP as compared with, for example, heat capacity
measurements \cite{18}. However, it is still a viable tool to identify
relative changes in the spectra caused by modification of the system
structure. Our Raman spectra results are discussed in conjunction with the
microstructure information based on X-ray and electron-diffraction experiments
made on these samples. In particular, our study illustrates how various
aspects of structural disorder affect electronic properties such as
conductivity as compared with their effect on the BP shape and magnitude. The
similarities and differences with the behavior of the BP in other glasses were
pressure was used to modify the structure are pointed out.

\section{Experimental}

\subsection{Samples preparation and characterization}

The In$_{\text{x}}$O films were e-gun evaporated onto room-temperature
substrates using 99.999\% pure In$_{\text{2}}$O$_{\text{3}}$
sputtering-target. Deposition was carried out at the ambience of 3$\pm
$0.5x10$^{\text{-5}}$ to 4$\pm$0.5x10$^{\text{-4}}$ Torr oxygen-pressure
maintained by leaking 99.9\% pure O$_{\text{2}}$ through a needle valve into
the vacuum chamber (base pressure $\simeq$10$^{\text{-6}}$ Torr). Different
substrates were used for the samples prepared for the different measurements
techniques. Undoped silicon wafers were used as substrates for electrical
measurements, X-ray diffraction (XRD) and Raman spectroscopy measurements.
Carbon-coated copper grids were used for transmission electron microscopy
(TEM) imaging and electron-diffraction. During deposition and
thermal-treatment the grids were anchored to glass-slides by small indium
balls pressed onto the glass. The deposited film on the rest of the slide was
used for monitoring the sample resistance. X-ray reflectometry (XRR)
measurements were performed on samples deposited on 3.8 mm thick float-glass-slides.

Rates of deposition in the range 0.3-2.5~\AA /s were used to produce films
with different compositions; The In$_{\text{x}}$O samples had
carrier-concentration \textit{N }that increases with the ratio of
deposition-rate to the oxygen-partial-pressure. For the rates-pressures used
here \textit{N} was in the range 2x10$^{\text{19}}$cm$^{\text{-3}}$ to
5x10$^{\text{21}}$cm$^{\text{-3}}$ as measured by Hall effect at
room-temperature using a Hall-bar control-sample prepared simultaneously for
each sample deposition. The evaporation source to substrate distance in the
deposition chamber was 45$\pm$1cm. This yielded films with thickness
uniformity of $\pm$2\% across a 2x2cm$^{\text{2}}$ area. Lateral sizes of
samples used for transport measurements was typically 1x2mm$^{\text{2}}$
(width x length respectively), and 1x1cm$^{\text{2}}$ for the Raman
spectroscopy. To afford reasonable resolution for electron-microscopy
thickness of the films used for TEM work was typically d=200$\pm$10 \AA .

Three batches of In$_{\text{x}}$O with different carrier-concentrations
\textit{N=(}4$\pm$1)x10$^{\text{19}}$cm$^{\text{-3}}$, (1$\pm$%
0.5)x10$^{\text{20}}$cm$^{\text{-3}}$, and (9$\pm$1)x10$^{\text{20}}%
$cm$^{\text{-3}}$ were used for Raman spectroscopy. For structural analysis we
used films characterized by the Ioffe-Regel parameter k$_{\text{F}}\ell
$=(3$\pi^{\text{2}}$)$^{\text{2/3}}\frac{\hbar\sigma_{\text{RT}}}{e^{\text{2}%
}\mathit{N}^{\text{1/3}}}$ in the range of 0.08 to 0.4. Here $\sigma
_{\text{RT}}$ is the room-temperature conductivity. This range covers the
critical-value of k$_{\text{F}}\ell$=0.32$\pm$0.2 where the metal-insulator
transition and superconductor-insulator transition of In$_{\text{x}}$O take
place \cite{19,20}.

High resolution TEM images and electron diffraction patterns were taken with
the Philips Tecnai F20 G2 operating at 200kV. X-ray diffraction and
reflectometry were taken with Bruker diffractometer AXS D8 Advance equipped
with Lynexeye XE-T silicon strip-detector. The diffractometer has step
resolution of 10$^{\text{-4}}$deg.

Raman spectroscopy at Ben-Gurion University of the Negev was performed with a
home-built ultra-low frequency Raman microscope confocal system that was
assembled and optimized for measurements down to 10cm$^{\text{-1}}$. The setup
employed a single longitudinal mode green-laser operating at 532nm. The laser
power used in the measurements was typically 3mW for a spot diameter of
$\approx$2.6$\mu$m. The scattered signal was collected via high throughput 532
longpass nano-edge filters and a single spectrometer \cite{21}. Complementary
Raman spectroscopy studies in the Hebrew University were taken with a Renishaw
inVia Reflex Spectrometer using a laser beam with either 514 nm or 785 nm
wavelength and edge-filter at $\approx$70cm$^{\text{-1}}$ therefore these
measurements were limited to energies $\gtrsim$80cm$^{\text{-1}}$.

\subsection{The thermal-treatment protocol}

The protocol we routinely use for monitoring the annealing process involves
the following steps: After removal from the deposition chamber, the sample was
mounted onto a heat-stage in a small vacuum-cell wired to make contacts with
the sample for electrical measurements, and a thermocouple attached to the
sample-stage as a thermometer. Resistance measurements were performed by a
two-terminal technique using either the computer-controlled HP34410A
multimeter or the Keithley K617. Next the heating stage is energized, and the
resistance and temperature is continuously measured throughout the heating,
relaxation, and cooling periods. A typical annealing cycle is illustrated in
figure 1:%

%TCIMACRO{\FRAME{ftbpFU}{3.4411in}{2.2857in}{0pt}{\Qcb{A typical protocol used
%in thermal-treating In$_{\text{x}}$O films. Resistance data R(t) are shown in
%squares and refer to the left scale, the sample temperature above room's
%$\Delta$T(t)$\equiv$T(heater)-T(room-temperature) is plotted with circles and
%refer to the right scale. The sample here has \QTR{it}{N}$\QTR{it}{\approx
%}$8.5x10$^{\text{20}}$cm$^{\text{-3}}$, thickness of 52nm and lateral
%dimension of 1x1mm$^{\text{2}}$. }}{}{fig_1.eps}%
%width 3.4411in;  height 2.2857in;  depth 0pt;  original-width 11.1578in;
%original-height 7.3812in;  cropleft "0";  croptop "1";  cropright "1";
%cropbottom "0";  filename 'Fig_1.eps';file-properties "XNPEU";}} }%
%BeginExpansion
\begin{figure}[ptb]%
\centering
\includegraphics[
height=2.2857in,
width=3.4411in
]%
{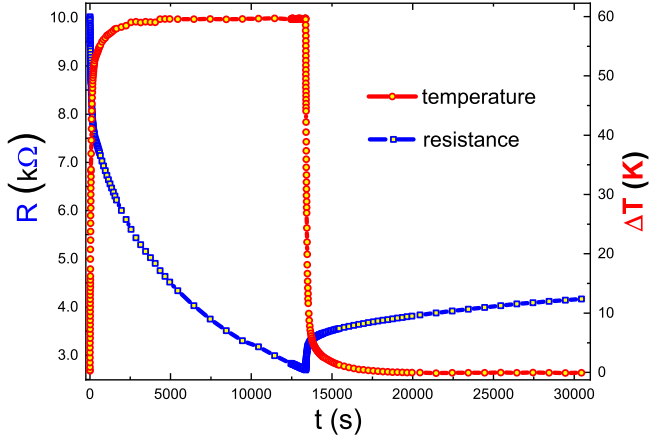}%
\caption{A typical protocol used in thermal-treating In$_{\text{x}}$O films.
Resistance data R(t) are shown in squares and refer to the left scale, the
sample temperature above room's $\Delta$T(t)$\equiv$%
T(heater)-T(room-temperature) is plotted with circles and refer to the right
scale. The sample here has \textit{N}$\mathit{\approx}$8.5x10$^{\text{20}}%
$cm$^{\text{-3}}$, thickness of 52nm and lateral dimension of 1x1mm$^{\text{2}%
}$. }%
\end{figure}
%EndExpansion
The sharp changes in the sample resistance R when the heating is turned on and
off are mostly due to the temperature dependent coefficient that in this
disorder regime is negative. However, during the time that the temperature has
settled at the target value for annealing, R continues to slowly decrease.
Then, after cooldown to room-temperature, the resistance slowly increases
approaching an asymptotic value that, ordinarily, is smaller than at the start
of the heat-treatment cycle.

These slow $\Delta$R(t) reflect changes in the system volume (densification
and rarefaction respectively). The change in volume in the process of
thermally-treating In$_{\text{x}}$O films was demonstrated in an interference
experiment using grazing-angle X-ray technique \cite{22}. It was further
correlated with in-situ resistivity and optical-spectroscopy measurements
\cite{22}. For In$_{\text{x}}$O, a system with the Fermi energy in the
conduction band, higher density typically yields higher mobility. This may be
due to the enhanced overlap of wavefunctions and improved connectivity. It
will be shown below that there is more to the improved mobility than just densification.

The time-dependent processes that occur while the temperature is constant were
qualitatively accounted for by a heuristic model based on the
two-level-systems that make up the potential landscape of the disordered
system \cite{23}. A similar approach was used in \cite{24} to offer a platform
for accounting for thermal expansion of glasses.

\section{Results and discussion}

\subsection{Raman spectra}

Figure 2 shows the measured Raman spectra of three batches of In$_{\text{x}}$O
films before and after thermal-treatment. The samples differ by their O-In
ratio determined during the deposition process. They are identified in the
figure by their carrier-concentration \textit{N} measured by the Hall effect.
These three compositions were chosen to represent the high-\textit{N},
medium-\textit{N} and low-\textit{N} versions of In$_{\text{x}}$O. The
high-\textit{N} and low-\textit{N} versions of In$_{\text{x}}$O in particular
exhibit different behavior in transport \cite{20}, and as will be shown below,
they differ in terms of other material properties.
%TCIMACRO{\FRAME{ftbpFU}{3.4411in}{6.397in}{0pt}{\Qcb{The Raman spectra for the
%three studied batches of In$_{\text{x}}$O measured down to $\approx
%$10cm$^{\text{-1}}$ with a laser wavelength $\lambda$=532nm. Spectra are shown
%before and after heat-treatment. The Ioffe-Regel parameters k$_{\text{F}}\ell$
%for these samples are: (a) before-0.11; after-0.39 (b) before-0.12; after-0.41
%(c) before-0.078; after-0.42. Blue curves are scaled-up copies of the
%heat-treated spectra (the factor is chosen to match the intensity reading at
%the peak). Note that the heat-treated curve shows a wider high-energy tail
%than the as-made samples of both sample (a) and sample (b).}}{}{fig_2.eps}%
%width 3.4411in;  height 6.397in;  depth 0pt;  original-width 9.1376in;
%original-height 6.8242in;  cropleft "0";  croptop "1";  cropright "1";
%cropbottom "0";  filename 'Fig_2.eps';file-properties "XNPEU";}} }%
%BeginExpansion
\begin{figure}[ptb]%
\centering
\includegraphics[
height=6.397in,
width=3.4411in
]%
{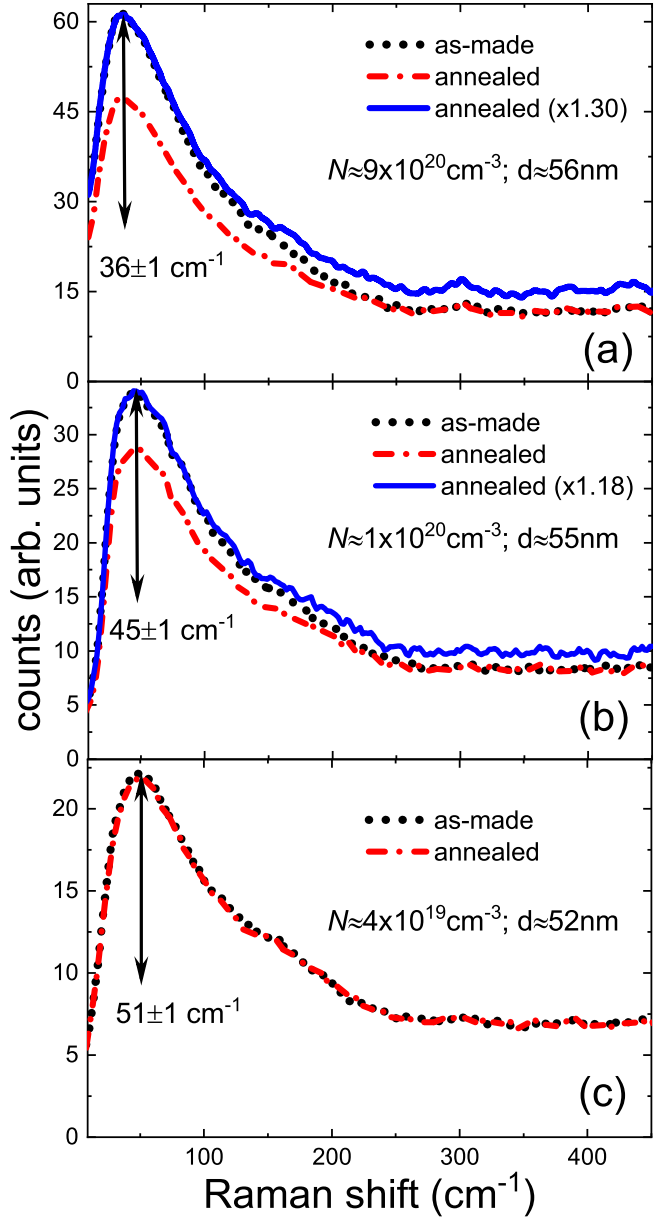}%
\caption{The Raman spectra for the three studied batches of In$_{\text{x}}$O
measured down to $\approx$10cm$^{\text{-1}}$ with a laser wavelength $\lambda
$=532nm. Spectra are shown before and after heat-treatment. The Ioffe-Regel
parameters k$_{\text{F}}\ell$ for these samples are: (a) before-0.11;
after-0.39 (b) before-0.12; after-0.41 (c) before-0.078; after-0.42. Blue
curves are scaled-up copies of the heat-treated spectra (the factor is chosen
to match the intensity reading at the peak). Note that the heat-treated curve
shows a wider high-energy tail than the as-made samples of both sample (a) and
sample (b).}%
\end{figure}
%EndExpansion
To cater for the spatial intensity-variations of the scattered signal the
Raman spectra were normalized to the intensity of the as-made sample evaluated
at 450cm$^{\text{-1}}$ (averaged value over the interval 440-450cm$^{\text{-1}%
}$). This will be referred to as a "background" intensity I$_{\text{0}}$. The
ratio I$_{\text{P}}$/I$_{\text{0}}$, where I$_{\text{P}}$is the intensity at
the BP maximum, turns out to be a meaningful measure of the BP magnitude;
I$_{\text{0}}$ taken at different points across a given specimen may vary by
more than $\approx$30\% while I$_{\text{P}}$/I$_{\text{0}}$ appears to be
constant to better than 2\% \cite{25}.

Qualitatively, the Raman spectra for the three batches in Fig.2 exhibit the
same BP shape characteristic of other amorphous and glassy systems
\cite{3,4,5,6,7,8,9,10,11,12,13,14,15,16}. There are however two quantitative
differences depending on the composition of the material. First, the peak
position increases with the O-In ratio (Fig.3).
%TCIMACRO{\FRAME{ftbpFU}{3.4411in}{2.4699in}{0pt}{\Qcb{The dependence of the
%energy of the BP maximum $\omega_{_{\text{BP}}}$ on the In$_{\text{x}}$O
%composition O/In. This is obtained through the relation between \QTR{it}{N}
%and the oxygen/indium mass ratio studied elsewhere (Fig.6 in \cite{19}).}}%
%{}{fig_3.eps}{\special{ language "Scientific Word";  type "GRAPHIC";
%width 3.4411in;  height 2.4699in;  depth 0pt;  original-width 14.2305in;
%original-height 28.5276in;  cropleft "0";  croptop "1";  cropright "1";
%cropbottom "0";  filename 'Fig_3.eps';file-properties "XNPEU";}} }%
%BeginExpansion
\begin{figure}[ptb]%
\centering
\includegraphics[
height=2.4699in,
width=3.4411in
]%
{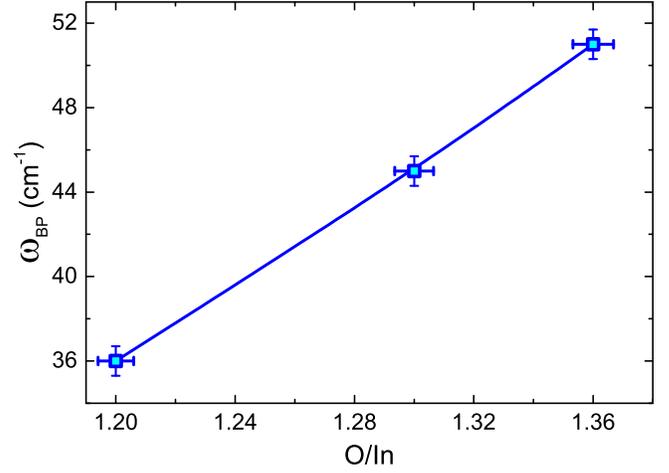}%
\caption{The dependence of the energy of the BP maximum $\omega_{_{\text{BP}}%
}$ on the In$_{\text{x}}$O composition O/In. This is obtained through the
relation between \textit{N} and the oxygen/indium mass ratio studied elsewhere
(Fig.6 in \cite{19}).}%
\end{figure}
%EndExpansion
The position of the BP scales with the typical phonon energy of the material
which is naturally smaller for the In-richer In$_{\text{x}}$O so this is just
a consequence of the batch composition. That the peak-position appears at a
frequency $\omega_{\text{p}}$(O/In) and does not shift due to the
thermal-treatment is consistent with the finding that the Hall coefficient is
unchanged in the process \cite{2}.

Secondly, the heat-treatment causes a more conspicuous decrease of the BP
magnitude for the higher-\textit{N }version while having indistinguishable
change in the spectra of the low-\textit{N} version despite the comparable
change in the samples resistance during annealing. Also, for a comparable
conductivity, the relative magnitude of the BP, is larger the higher is the carrier-concentration.

The amount of disorder required to affect a given change of the conductivity
(or k$_{\text{F}}\ell$) in a degenerate Fermi system like In$_{\text{x}}$O
grows with the Fermi-energy (thus with \textit{N}). This correlation was
demonstrated in the dependence of the optical gap on k$_{\text{F}}\ell$
studied in-situ in \cite{2}. Figure 4 shows three curves from this study for
samples with carrier-concentration that are close to these of the batches
studied in Fig.2. Note that, for a comparable change of k$_{\text{F}}\ell$ a
larger change of the optical-gap occurred for the sample with the higher
carrier-concentration, a similar trend to that observed in the dependence of
the BP on disorder (Fig.2). The relative change of the BP magnitude presumably
reflects the degree of the structural-change that occurred during
heat-treatment. The correlation between disorder (either chemical or
electronic) and I$_{\text{P}}$/I$_{\text{0}}$ (table 1) has the same reason -
for a comparable k$_{\text{F}}\ell$ higher disorder yields a more conspicuous
BP demonstrating the common observation related to the phenomenon \cite{12}.%
%TCIMACRO{\FRAME{ftbpFU}{3.4411in}{4.5973in}{0pt}{\Qcb{The dependence of the
%optical gap on the Ioffe-Regel parameter k$_{\text{F}}\ell$ for different
%versions of In$_{\text{x}}$O labeled by their carrier-concentration. Note that
%a larger change of E$_{\text{g}}$ is required to affect a given change in
%k$_{\text{F}}\ell$ the larger \QTR{it}{N} is. The arrows mark the value of
%k$_{\text{F}}\ell$ for the as-made and annealed states of the three samples of
%Fig.2; Red -Fig.2a, Green-Fig.2b, Blue-Fig.2c.}}{}{fig_4.eps}%
%width 3.4411in;  height 4.5973in;  depth 0pt;  original-width 7.5213in;
%original-height 10.0707in;  cropleft "0";  croptop "1";  cropright "1";
%cropbottom "0";  filename 'Fig_4.eps';file-properties "XNPEU";}} }%
%BeginExpansion
\begin{figure}[ptb]%
\centering
\includegraphics[
height=4.5973in,
width=3.4411in
]%
{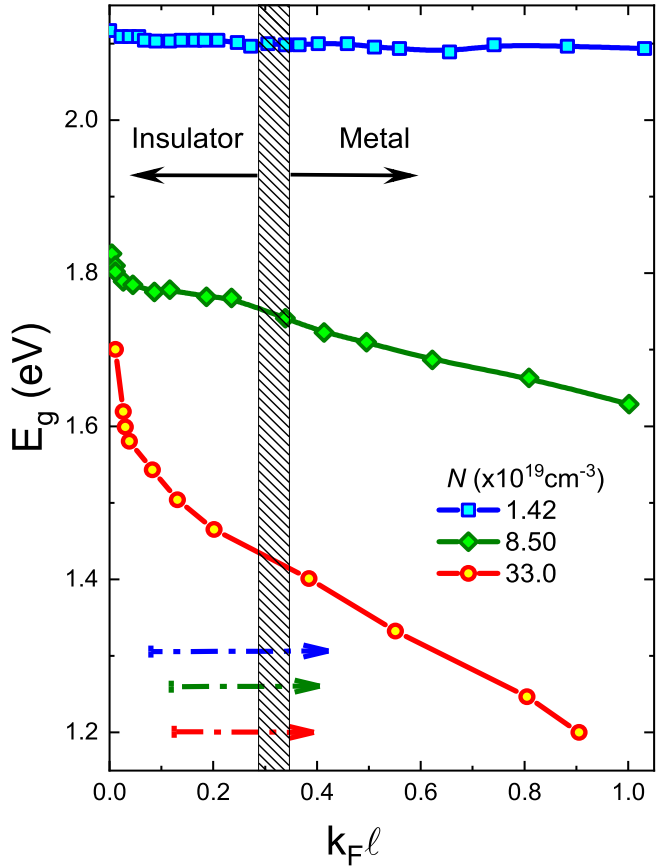}%
\caption{The dependence of the optical gap on the Ioffe-Regel parameter
k$_{\text{F}}\ell$ for different versions of In$_{\text{x}}$O labeled by their
carrier-concentration. Note that a larger change of E$_{\text{g}}$ is required
to affect a given change in k$_{\text{F}}\ell$ the larger \textit{N} is. The
arrows mark the value of k$_{\text{F}}\ell$ for the as-made and annealed
states of the three samples of Fig.2; Red -Fig.2a, Green-Fig.2b, Blue-Fig.2c.}%
\end{figure}
%EndExpansion

Table 1 includes the relative magnitude of the BP before and after annealing,
with the batch chemical-disorder $\delta$[O/In], and a quantitative measure of
disorder W$_{\text{C}}$ based on data of electronic-transport. The chemical
disorder is taken here as the deviation of the composition from that of the
stoichiometric In$_{\text{2}}$O$_{\text{3}}$ compound. W$_{\text{C}}$ is the
critical disorder that Anderson-localizes the particular batch. W$_{\text{C}%
}\propto$E$_{\text{F}}\propto$\textit{N}$^{\text{2/3}}$ where E$_{\text{F}}$,
is the Fermi energy. For In$_{\text{x}}$O the proportionality factor between
disorder and Fermi-energy was found to be: W$_{\text{C}}\simeq$%
6.2\textperiodcentered E$_{\text{F}}$ \cite{2} and E$_{\text{F}}$ of a given
batch is obtained using free-electron formulae.

\begin{center}%
\begin{tabular}
[c]{|c|c|c|c|}\hline
$\delta$\textbf{[O/In]} & \textbf{W}$_{\text{\textbf{C}}}$\textbf{ (eV)} &
\textbf{I}$_{\text{P}}$\textbf{/I}$_{\text{0}}$\textbf{ as-made} &
\textbf{I}$_{\text{P}}$\textbf{/I}$_{\text{0}}$\textbf{ annealed}\\\hline
0.3 & 2.21 & 5.3 & 4.2\\\hline
0.2 & 0.51 & 4.1 & 3.4\\\hline
0.14 & 0.28 & 3.1 & 3.1\\\hline
\end{tabular}
\newline
\end{center}

{\small Table 1: Values of parameters for the three In}$_{\text{x}}${\small O
batches measured in Fig.2. The electronic disorder (characterized by
k}$_{\text{F}}\ell${\small ) includes the contribution of the deviation from
stoichiometry that may is relevant for phonon scattering vs. electron
scattering discussed in the text below. Note the systematic dependence of the
BP magnitude on the batch disorder.}

\subsection{Structural changes resulting from the heat-treatment}

Before proceeding with further analysis of the Raman spectra we digress now to
see what actually changes in the system micro-structure due to heat-treatment.
This was done by using customary tools of structural analysis; X-ray and
electron microscopy. Special emphasis was given to the high-\textit{N }version
of In$_{\text{x}}$O where the effect in terms of Raman spectroscopy is
manifestly large. Consider first the electron-diffraction and TEM images for a
typical sample shown in Fig.5a and Fig.5b:%
%TCIMACRO{\FRAME{ftbpFU}{3.4411in}{3.4774in}{0pt}{\Qcb{ Bright-field TEM images
%and the associated diffraction patterns for a 20nm thick In$_{\text{x}}$O
%sample deposited on carbon coated Cu grid..(a) as-made and (b) the same sample
%after heat-treatment. The control sample, deposited on a glass-substrate, had
%a k$_{\text{F}}\ell$ of 0.16 and 0.39 before and after treatment
%respectively.}}{}{fig_5.eps}{\special{ language "Scientific Word";
%type "GRAPHIC";  maintain-aspect-ratio TRUE;  display "USEDEF";
%valid_file "F";  width 3.4411in;  height 3.4774in;  depth 0pt;
%original-width 14.0679in;  original-height 14.2227in;  cropleft "0";
%filename 'Fig_5.eps';file-properties "XNPEU";}} }%
%BeginExpansion
\begin{figure}[ptb]%
\centering
\includegraphics[
height=3.4774in,
width=3.4411in
]%
{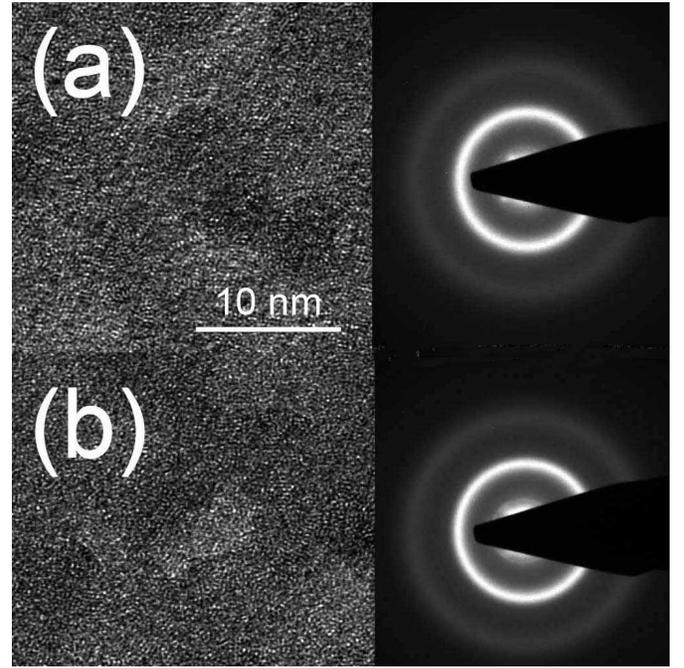}%
\caption{ Bright-field TEM images and the associated diffraction patterns for
a 20nm thick In$_{\text{x}}$O sample deposited on carbon coated Cu grid..(a)
as-made and (b) the same sample after heat-treatment. The control sample,
deposited on a glass-substrate, had a k$_{\text{F}}\ell$ of 0.16 and 0.39
before and after treatment respectively.}%
\end{figure}
%EndExpansion
Both diffraction patterns exhibit broad rings characteristic of amorphous
structure with no sign of crystallization. In fact, it is hard to see
difference in the before and after patterns. On closer examination, the first
strong ring in the pattern is sharper in the annealed sample and the
associated bright-field image appears somewhat softer.
%TCIMACRO{\FRAME{ftbpFU}{3.4411in}{3.2188in}{0pt}{\Qcb{Intensity profile for
%the two diffraction patterns shown in Fig.5. Top: Scanned across the first
%strong ring diameter D. Bottom: The profile of the first strong ring clearly
%showing the change in the width $\Gamma$ after heat-treatment.}}{}%
%{fig_6.eps}{\special{ language "Scientific Word";  type "GRAPHIC";
%width 3.4411in;  height 3.2188in;  depth 0pt;  original-width 9.2907in;
%original-height 8.6896in;  cropleft "0";  croptop "1";  cropright "1";
%cropbottom "0";  filename 'Fig_6.eps';file-properties "XNPEU";}} }%
%BeginExpansion
\begin{figure}[ptb]%
\centering
\includegraphics[
height=3.2188in,
width=3.4411in
]%
{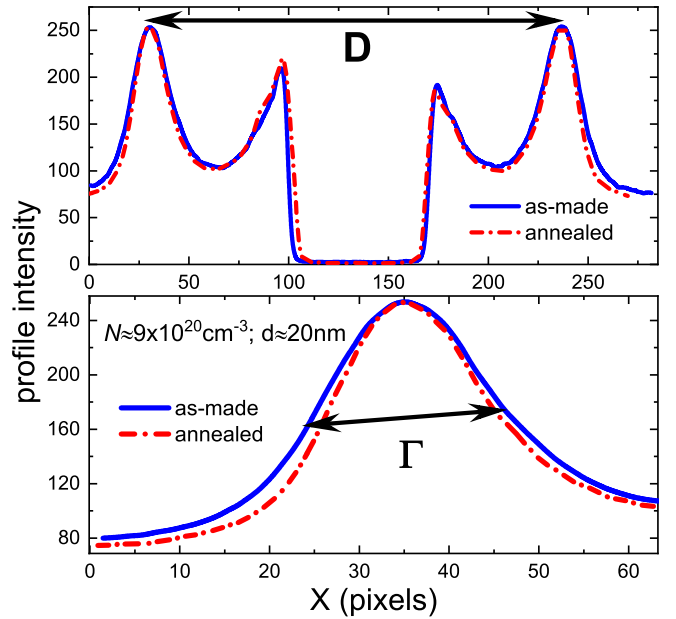}%
\caption{Intensity profile for the two diffraction patterns shown in Fig.5.
Top: Scanned across the first strong ring diameter D. Bottom: The profile of
the first strong ring clearly showing the change in the width $\Gamma$ after
heat-treatment.}%
\end{figure}
%EndExpansion

The changes in the diffraction pattern may be quantified by recording the
intensity profiles of the diffraction patterns as is illustrated in Figure 6.
The measurement confirms the eye-impression; the width of the ring $\Gamma$
decreased by $\approx$12\% in the annealed sample. At the same time, the
average ring-diameter indicated by D in the figure, \textit{increased} by
$\approx$0.4$\pm$0.1\% suggesting a reduced interatomic separation due to annealing.

Similar results were obtained from XRD measurements on this high-\textit{N
}sample as shown in Fig.7.%
%TCIMACRO{\FRAME{ftbpFU}{3.4411in}{3.0727in}{0pt}{\Qcb{X-ray diffraction taken
%over the first strong diffraction ring of the sample that had k$_{\text{F}%
%}\ell$=0.09 and k$_{\text{F}}\ell$=0.38 for as-made and annealed samples
%respectively. (a) Are the raw data, and the dashed and dotted lines stand for
%the background intensities. These background lines are subtracted from the raw
%data and fitted in (b) to: A\textperiodcentered exp$\left[
%\text{-0.5\textperiodcentered}\left(  \frac{\text{X-X}_{_{\text{0}}}}{\sigma
%}\right)  ^{\text{2}}\right]  $ where A is the intensity amplitude and
%X$\equiv$2$\theta$. The fits, shown as dashed lines, yield X$_{\text{0}}%
%$=31.78$\pm$0.008; $\sigma$=2.4$\pm$0.01 and X$_{\text{0}}$=31.98$\pm$0.008;
%$\sigma$=2.2$\pm$0.01 for the as-made and annealed plots respectively. The
%larger intensity at the peak (bottom graph) is due to the reduced background
%and the narrower line width.}}{}{fig_7.eps}%
%width 3.4411in;  height 3.0727in;  depth 0pt;  original-width 9.8338in;
%original-height 8.7744in;  cropleft "0";  croptop "1";  cropright "1";
%cropbottom "0";  filename 'Fig_7.eps';file-properties "XNPEU";}} }%
%BeginExpansion
\begin{figure}[ptb]%
\centering
\includegraphics[
height=3.0727in,
width=3.4411in
]%
{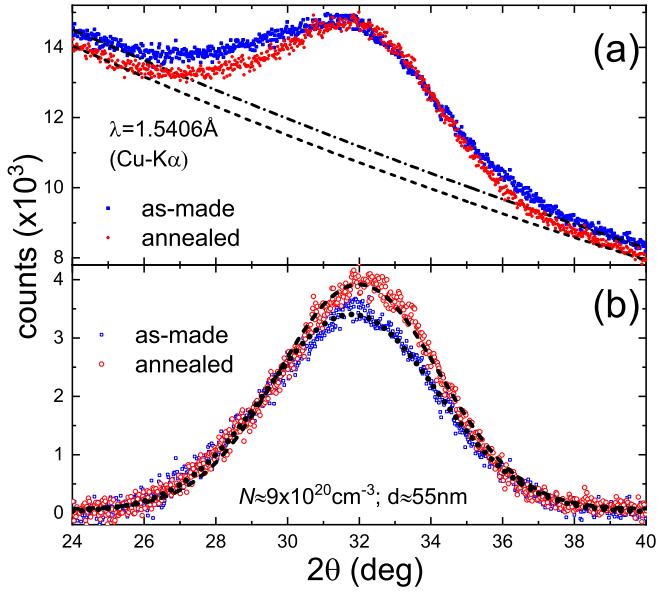}%
\caption{X-ray diffraction taken over the first strong diffraction ring of the
sample that had k$_{\text{F}}\ell$=0.09 and k$_{\text{F}}\ell$=0.38 for
as-made and annealed samples respectively. (a) Are the raw data, and the
dashed and dotted lines stand for the background intensities. These background
lines are subtracted from the raw data and fitted in (b) to:
A\textperiodcentered exp$\left[  \text{-0.5\textperiodcentered}\left(
\frac{\text{X-X}_{_{\text{0}}}}{\sigma}\right)  ^{\text{2}}\right]  $ where A
is the intensity amplitude and X$\equiv$2$\theta$. The fits, shown as dashed
lines, yield X$_{\text{0}}$=31.78$\pm$0.008; $\sigma$=2.4$\pm$0.01 and
X$_{\text{0}}$=31.98$\pm$0.008; $\sigma$=2.2$\pm$0.01 for the as-made and
annealed plots respectively. The larger intensity at the peak (bottom graph)
is due to the reduced background and the narrower line width.}%
\end{figure}
%EndExpansion
The small differences between the XRD and electron-diffraction in terms of the
changes in D and ring-width may be due to the different substrates used.

On the basis of these measurements one might conclude that the volume-change
-$\Delta$V/V of this sample due to the thermal-treatment is of the order of
1.2\% to 1.8\%. However, the results of XRR measurement suggest that $\Delta
$V/V for this sample may be significantly larger (Fig.8a); Following treatment
the sample thickness was reduced by $\approx$3.3\% implying a volume-change
-$\Delta$V/V of the order of $\approx$10\%. Similar -$\Delta$V/V values during
heat-treatment were obtained in a previous study of In$_{\text{x}}$O \cite{2}.
The difference in -$\Delta$V/V derived from the XRD versus that of the XRR
suggest that the In$_{\text{x}}$O structure is made-up of loosely packed
aggregates of relatively dense material. Such a porous medium is common in
vapor-deposited films and more generally in substances that were quench-cooled
from high temperatures. Actually, porosity is an abundant property of many
materials. An extreme example of such a structure is a cotton-ball or a bundle
of steel-wool. The volume of these substances may be greatly reduced when
pressed while their solid part remains essentially intact.
%TCIMACRO{\FRAME{ftbpFU}{3.4402in}{3.9531in}{0pt}{\Qcb{X-ray reflectometry for
%two versions of In$_{\text{x}}$O films having similar composition as that of
%the samples in Fig.2. Plots are shown for the before-and-after heat-treatment
%(curves are displaced along the ordinate for clarity). The thickness of the
%sample in (a) changed during heat-treatment from 56.2$\pm$0.1nm to 54.5$\pm
%$0.1nm and the sample in (b) changed from 45.1$\pm$0.1nm to 45$\pm$0.1nm. The
%films roughness parameters are: 1$\pm$0.2 for sample in (a) and 0.4$\pm$0.1
%for sample in (b).}}{}{fig_8.eps}{\special{ language "Scientific Word";
%type "GRAPHIC";  maintain-aspect-ratio TRUE;  display "USEDEF";
%valid_file "F";  width 3.4402in;  height 3.9531in;  depth 0pt;
%original-width 9.5813in;  original-height 10.1131in;  cropleft "0";
%filename 'Fig_8.eps';file-properties "XNPEU";}} }%
%BeginExpansion
\begin{figure}[ptb]%
\centering
\includegraphics[
height=3.9531in,
width=3.4402in
]%
{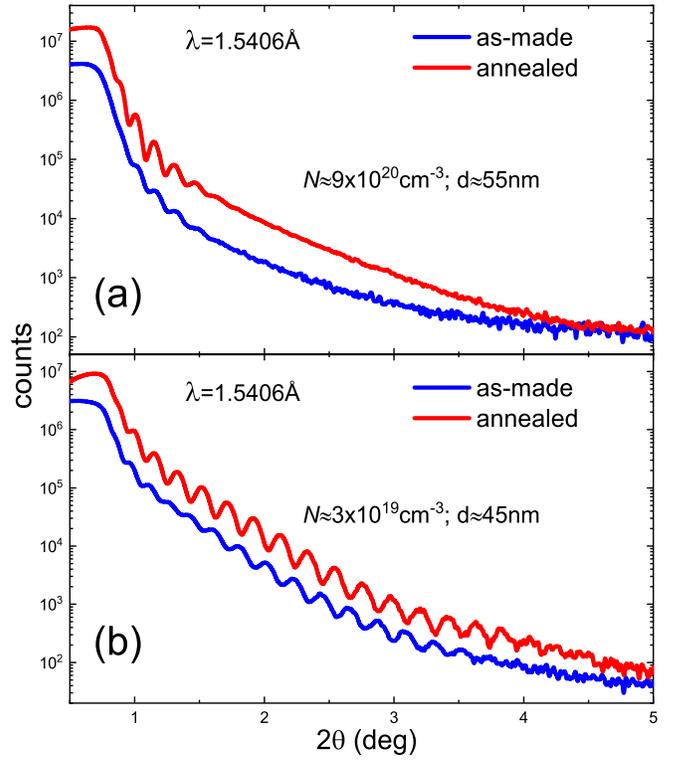}%
\caption{X-ray reflectometry for two versions of In$_{\text{x}}$O films having
similar composition as that of the samples in Fig.2. Plots are shown for the
before-and-after heat-treatment (curves are displaced along the ordinate for
clarity). The thickness of the sample in (a) changed during heat-treatment
from 56.2$\pm$0.1nm to 54.5$\pm$0.1nm and the sample in (b) changed from
45.1$\pm$0.1nm to 45$\pm$0.1nm. The films roughness parameters are: 1$\pm$0.2
for sample in (a) and 0.4$\pm$0.1 for sample in (b).}%
\end{figure}
%EndExpansion

Figure 8b shows the XRR trace taken of a low-\textit{N} sample that was added
for comparison. It exhibits a quantitatively different behavior than the
high-\textit{N} sample in two aspects; the visibility of the interference
extends over a wider range of angles, and the change of thickness during
heat-treatment is much smaller (per the same change of the sample resistance).
Note that heat-treatment enhanced the interference visibility in both samples
yet the low-\textit{N} sample retains a smoother film surface even when its
resistance is considerably higher. Evidently, for a similar k$_{\text{F}}\ell
$, the high-\textit{N} version of In$_{\text{x}}$O is both, more disordered
and has a rougher surface.

To understand how structural aspects affect conductivity and the BP it may be
useful to review the specific ingredients responsible for scattering in
In$_{\text{x}}$O films.

\subsection{The elements of disorder in amorphous indium-oxide films and their
effect on conductivity and the boson-peak}

As amorphous material In$_{\text{x}}$O lacks long-range order which by itself
imposes limit on charge mobility. The vacuum deposited In$_{\text{x}}$O films
have ancillary sources of disorder that lead to scattering and restrict their
k$_{\text{F}}\ell$ value. First, there is an off-diagonal disorder in the
material that is related to the distributed nature of the inter-atomic
separation. The distribution of inter-particle distances gets narrower as the
volume decreases and the system approaches the `ideal' closed-packed amorphous
structure. This enhances wavefunction overlap and therefore it naturally
affects the conductivity. The diminishment of off-diagonal disorder during
heat-treatment is clearly reflected in the reduced width of the
electron-diffraction and in the XRD patterns (Figs.5,6 and Fig.7
respectively). The latter has been often associated with the appearance of
medium-range order in the system \cite{7}. This re-structure process is also
reflected in the rate-distribution of the glass dynamics \cite{23}. The
`free-volume' of the sample that is presumably eliminated is probably the most
important single element in giving rise to phonon scattering. In fact, a
porous nature of amorphous structures has been shown by detailed simulation
studies to be the major contributing factor in the BP magnitude \cite{12}.

A clear correlation between a structural-change and a modified BP is
manifested in our study whenever a significant change of density takes place.
This is observed most conspicuously for the heat-treated high-\textit{N}
sample where density is reduced by $\approx$10\% (Fig.8a) and the Raman
spectrum shows a $\approx$30\% reduction in magnitude (Fig.2a). The
correlation with the material density follows the trend reported in
experiments on other disordered systems where the magnitude of the BP was
observed to be smaller following densification by pressure
\cite{26,27,28,29,30,31,32,33,34,35}.

The main difference between heat-treatment and applying high pressure appears
to be the shift of the BP position to higher energy in the pressure
experiments \cite{26,27,28,29,30,31,32,33,34,35} whereas no such shift is
encountered in the Raman spectra for the thermally-annealed samples. In this
regard the evolution of the BP shape in our experiments followed the
heterogeneously-distributed elastic-constants scenario described by
Schirmacher and Ruocco \cite{8}. A similar behavior to our annealing
experiments was observed in the BP spectra of As$_{\text{2}}$S$_{\text{3}}$
samples after it was cold-quenched from a well-annealed state \cite{36}.

Another source of disorder in the amorphous indium-oxide system is associated
with deviation from stoichiometry - chemical disorder. Relative to the ionic
compound In$_{\text{2}}$O$_{\text{3-x}}$, there are 5-30\% oxygen vacancies in
In$_{\text{x}}$O spanning the range $\approx$10$^{\text{18}}$-5x10$^{\text{21}%
}$cm$^{\text{-3}}$ in terms of carrier-concentration{\small .} To preserve
chemical neutrality some indium atoms must assume a valence of +1 instead of
the +3 they have in the stoichiometric compound. When randomly distributed
this valence-fluctuation forms a background potential with an amplitude of the
order of few eV (assuming an average interatomic-separation of the order of
$\approx$0.3nm). This type of disorder is quite prevalent in
non-stoichiometric compounds, metallic-oxides, high-Tc materials etc. and it
seems to be a main source of elastic scattering in both In$_{\text{x}}$O and
In$_{\text{2}}$O$_{\text{3-x}}$ \cite{37}. On the other hand the role
chemical-disorder plays in the buildup of a BP is unclear. Deviation from
stoichiometry unaccompanied by other factors, does not necessarily promote
formation of a BP; Polycrystalline films of In$_{\text{2}}$O$_{\text{3-x}}$
exhibits 5-8\% oxygen vacancies \cite{37} while showing very small magnitude
of BP relative to the amorphous version \cite{38}. This probably means that
oxygen vacancies and larger pockets of free-volume such as di-vacancies, are
evenly distributed such that density-fluctuations over a phonon wavelength are
rather small.

Finally, the reduction of surface roughness during the heat-treatment revealed
in the XRR data needs elaboration. Changes in the interference visibility
presumably reflect re-arrangement of ions at the film surface. This seems to
occur even when changes in thickness were too small to be observed (see,
Fig.8b). In principle, a rough film surface is a source of scattering for both
phonons and electrons, and it is natural to expect less scattering when the
surface roughness is reduced. Given that the mean-free-path $\ell$ in our
samples is much smaller than the film thickness, the contribution of the
surface to scattering by either electrons or phonons is probably very small.
Electrical conductivity is sensitive enough to detect a small change of
disorder in the sample. For phonons however the same structural change may be
too small to affect the BP magnitude (Fig.2c). In other words, a measurable
change in conductivity may be affected without a significant structural
change, in similar vein with the dependence of the optical-gap E$_{\text{g}}$
on k$_{\text{F}}\ell$ for a low-\textit{N} sample where E$_{\text{g}}$ remains
almost constant while k$_{\text{F}}\ell$ changes over a large range (Fig.4).

\section{Summary}

We have followed by transport, structural-tools, and Raman spectroscopy the
changes that occur during heat-treating In$_{\text{x}}$O films. Transport
measurements were used to quantify the degree of disorder in In$_{\text{x}}$O
samples with different carrier-concentrations before and after treatment. The
disorder is characterized by assigning each sample a Ioffe-Regel parameter
k$_{\text{F}}\ell$. The study reveals a correlation between the system
disorder defined in this way, and the magnitude of the BP. This correlation
suggests that, in these systems, phonons are scattered by the same elements of
disorder that cause scattering of electrons although not necessarily with the
same efficiency.

An element of disorder that has a large effect on the BP magnitude is the
presence of `free-volume' in the system that, in In$_{\text{x}}$O is
presumably related to the spatial distribution of oxygen vacancies. These are
re-arranged during the annealing process to reduce the system volume, and the
BP magnitude is changed accordingly. Our study furnishes the experimental
support to the simulation work of Shintani and Tanaka that identified the most
conspicuous BP in low-density defective structures \cite{12}. This led them to
conclude: "...\textit{the origin of the boson peak (are) transverse
vibrational modes associated with low-density defective structures}" \cite{12}.

The emerging picture is that heat-treating In$_{\text{x}}$O is analogous to
the process of gently tapping a ground-coffee bag to pack it tighter. Tapping
supplies the energy necessary to overcome local barriers allowing the powder
to reduce its gravitational energy. Temperature and the
interparticle-attraction respectively play the analogous roles in the process
of densifying In$_{\text{x}}$O. Enhanced conductivity due to densification
follows from enhanced wavefunction overlap as well as from improved
connectivity. This is accompanied by a reduced disorder and therefore weaker
heterogeneity which is reflected in a smaller magnitude of the BP. The
flexibility that the In$_{\text{x}}$O system offers in terms of fine-tuning
disorder by heat-treatment makes it a prime candidate for the study of
electronic transport in glasses, and as demonstrated in this work, also for
other fundamental properties of amorphous materials.

\begin{acknowledgments}
We benefitted from discussions with Walter Schirmacher and Alessio Zaccone.
The assistance of Anna Radko, Vladimir Uvarov, and Inna Popov from the Center
for Nanoscience and Technology (HU) is gratefully acknowledged. This research
has been supported by the 1030/16 grant administered by the Israel Academy for
Sciences and Humanities.
\end{acknowledgments}

\section{Supplementary Material}

This part and the following subsections give auxiliary information related to
the methods and techniques employed in the study.

\subsection{Avoiding crystallization}

One should realize that the crystalline version of indium-oxide,
In$_{\text{2}}$O$_{\text{3-x}}$ differs markedly from these amorphous versions
that are investigated here. In$_{\text{x}}$O may be easily crystallized to
form polycrystalline In$_{\text{2}}$O$_{\text{3-x}}$ once exposed and
maintained at temperatures as low as 380K (depending on the In$_{\text{x}}$O
composition).The danger of In$_{\text{x}}$O crystallization must be taken into
account especially when exposing the film to intense laser radiation but also
during heat-treatment. Keeping the heat-treatment temperature below 370K
practically eliminates crystallization problems; we never encountered a
problem with heat-treatment even when the sample was left for 34 days at this
temperature. It is more complicated to set a limit for the laser power because
it depends on the heat-dissipation through the substrate. To assist in
identifying the threshold we carried an experiment designed to observe the
onset of structural changes using high-resolution TEM. The results are shown
in Fig.9 and Fig.10:%

%TCIMACRO{\FRAME{ftbpFU}{3.3399in}{2.3367in}{0pt}{\Qcb{Raman spectra taken
%consecutively on a single spot with different laser powers. The sample is a
%20nm In$_{\text{x}}$O film deposited on a carbon-coated TEM copper grid. Note
%the appearance of extra structure for the 0.7mW plot that gets further
%developed for the 0.14mW exposure. }}{}{fig_sm1.eps}%
%width 3.3399in;  height 2.3367in;  depth 0pt;  original-width 10.5715in;
%original-height 7.3682in;  cropleft "0";  croptop "1";  cropright "1";
%cropbottom "0";  filename 'Fig_SM1.eps';file-properties "XNPEU";}} }%
%BeginExpansion
\begin{figure}[ptb]%
\centering
\includegraphics[
height=2.3367in,
width=3.3399in
]%
{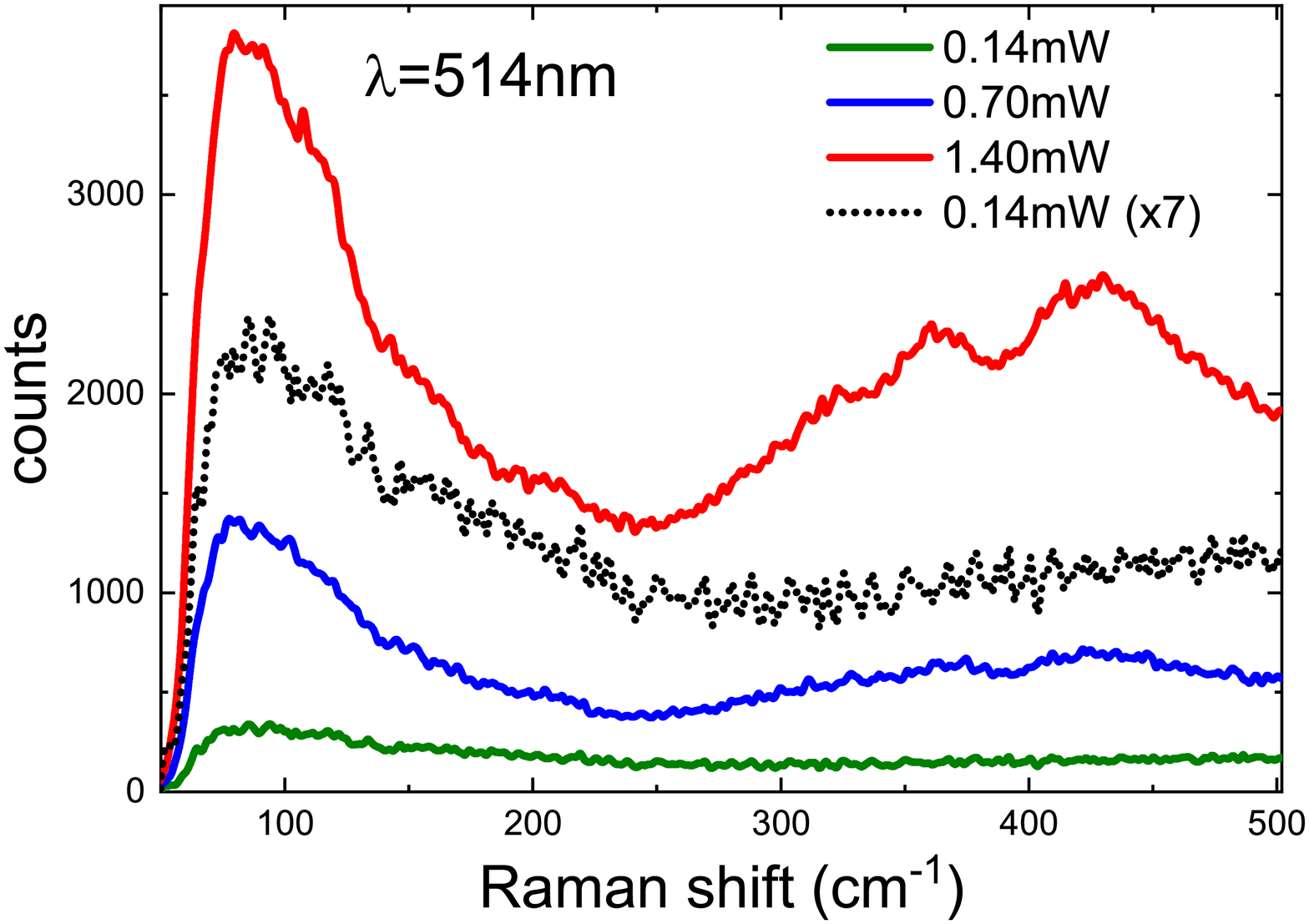}%
\caption{Raman spectra taken consecutively on a single spot with different
laser powers. The sample is a 20nm In$_{\text{x}}$O film deposited on a
carbon-coated TEM copper grid. Note the appearance of extra structure for the
0.7mW plot that gets further developed for the 0.14mW exposure. }%
\end{figure}
%EndExpansion
%

%TCIMACRO{\FRAME{ftbpFU}{3.3399in}{3.8112in}{0pt}{\Qcb{TEM and optical
%microscope pictures of the spot created by the laser exposures in SM1 above.
%The arrows mark the local area from which either bright-field image or
%diffraction pattern were taken. Note that the structural effect of the 2$\mu$m
%laser-spot extends over $\approx$10$\mu$m. The sample is a $\approx$2nm
%In$_{\text{x}}$O deposited on carbon-coated copper grid.}}{}{fig_sm2.eps}%
%width 3.3399in;  height 3.8112in;  depth 0pt;  original-width 7.0655in;
%original-height 8.0704in;  cropleft "0";  croptop "1";  cropright "1";
%cropbottom "0";  filename 'Fig_SM2.eps';file-properties "XNPEU";}} }%
%BeginExpansion
\begin{figure}[ptb]%
\centering
\includegraphics[
height=3.8112in,
width=3.3399in
]%
{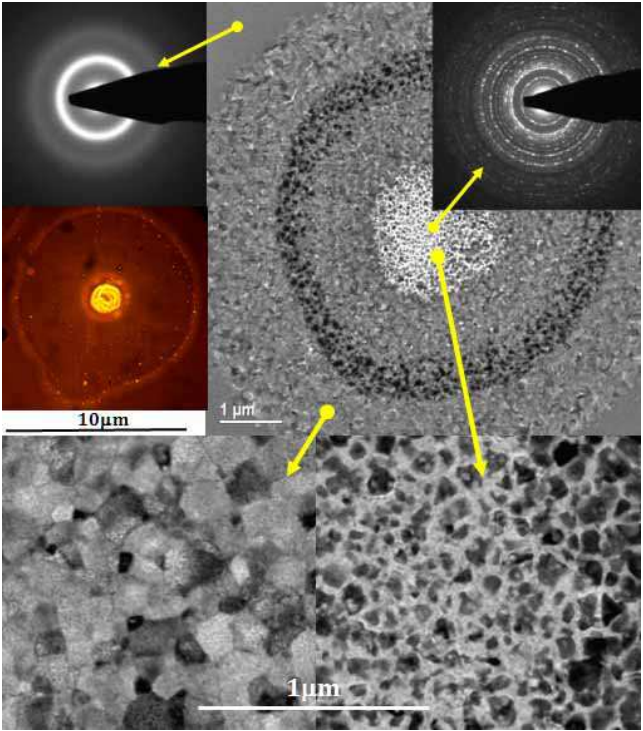}%
\caption{TEM and optical microscope pictures of the spot created by the laser
exposures in SM1 above. The arrows mark the local area from which either
bright-field image or diffraction pattern were taken. Note that the structural
effect of the 2$\mu$m laser-spot extends over $\approx$10$\mu$m. The sample is
a $\approx$2nm In$_{\text{x}}$O deposited on carbon-coated copper grid.}%
\end{figure}
%EndExpansion
On the basis of the correlation between these structural and Raman data we
identify the spectra taken with 0.7mW and 1.4mW laser power as a signature of
partial crystallization, or a highly defected polycrystalline phase (see
Fig.11 for typical spectrum of well-annealed In$_{\text{2}}$O$_{\text{3-x}}$
that lacks the pronounced BP observe in the laser-exposed plots in Fig.9).
This signature was used in the study to set the power-limit on samples
deposited on Si-wafers. The latter exhibit more efficient heat-dissipation
than the carbon film used for the TEM work allowing a larger laser power to be
used (see text).%

%TCIMACRO{\FRAME{ftbpFU}{3.3399in}{2.4033in}{0pt}{\Qcb{Raman spectra of two
%distinct structures of indium-oxide. The plot labeled as In$_{\text{2}}%
%$O$_{\text{3-x}}$ (red-line) is crystallized from the deposited amorphous film
%with thickness 90nm to form the polycrystalline version of indium-oxide. The
%blue dotted line depicts the spectrum taken from the deposited In$_{\text{x}}%
%$O film. The substarte is quartz.}}{}{fig_sm3.eps}%
%width 3.3399in;  height 2.4033in;  depth 0pt;  original-width 7.1581in;
%original-height 9.5683in;  cropleft "0";  croptop "1";  cropright "1";
%cropbottom "0";  filename 'Fig_SM3.eps';file-properties "XNPEU";}} }%
%BeginExpansion
\begin{figure}[ptb]%
\centering
\includegraphics[
height=2.4033in,
width=3.3399in
]%
{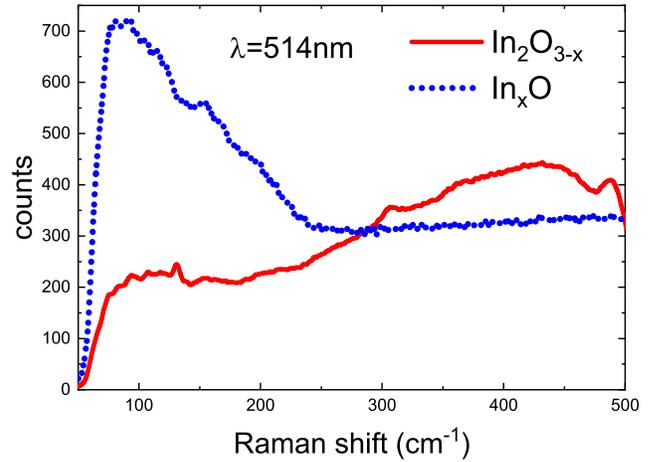}%
\caption{Raman spectra of two distinct structures of indium-oxide. The plot
labeled as In$_{\text{2}}$O$_{\text{3-x}}$ (red-line) is crystallized from the
deposited amorphous film with thickness 90nm to form the polycrystalline
version of indium-oxide. The blue dotted line depicts the spectrum taken from
the deposited In$_{\text{x}}$O film. The substarte is quartz.}%
\end{figure}
%EndExpansion

\subsection{Low-energy Raman setup}

A home-built ultra-low frequency Raman confocal system was assembled and
optimized for measurements of Raman spectra down to 10cm$^{\text{-1}}$. The
setup included an exciting-source Nd:YAG laser (CNI, MSL-FN-532), and the
scattered signal was collected via high throughput 532 longpass nano-edge
filters (NEFs) and a single spectrometer.

In particular, a single longitudal mode doubled Nd:YAG laser (CNI, MSL-FN-532)
beam from a diode-pumped solid-state laser was directed toward a reflecting
Optigrate band-pass filter, based on the volume Bragg-grating technique for
cleaning up the laser spectral noise and resulting in spectral widths
%TCIMACRO{\TEXTsymbol{<} }%
%BeginExpansion
$<$
%EndExpansion
7 cm$^{\text{-1}}$. Then two mirrors were used to aim the reflected laser beam
towards a 532 NEF (Iridian spectral technologies) with deep blocking (%
%TCIMACRO{\TEXTsymbol{>}}%
%BeginExpansion
$>$%
%EndExpansion
OD 6) at the laser line, set in an adjustable filter holder. The combination
of the mirrors with the NEF1, allowed to adjust the angle between the incident
laser beam and the normal of NEF1 for gaining the best laser line attenuation
and the lowest possible frequency of the ULF Raman signal, as well as to
reflect the beam to the microscope objective. The laser beam was focused
through a X50/0.65 microscope objective to provide a 3 mW incident beam on the
sample, which also collected and collimated the back scattered Rayleigh and
Raman signals. The backscattered signal passed through the above mentioned
NEF1, partially filtering the signal and then through an additional filter,
NEF2, to achieve the desirable attenuation of the Rayleigh signal. Following
the passage through the filters, the signal was focused by a 10 cm focal
length plano-convex lens onto a pinhole (100 $\mu$m), located in front of a 10
$\mu$m slit of a 0.14 m Czerny-Turner spectrometer (Jobin-Yvon, MicroHR) with
an entrance aperture ratio of f/3.88, with a 1200 g/mm grating. Finally, the
signal was detected by an air cooled 1,024 x 1,024 intensified charge-coupled
device (Andor, DH734-18U), driven by the Solis 4.3 software and analyzed. The
system can achieve spectral resolution at the full-width half-maximum of a
peak in the spectrum of up to 11 cm$^{\text{-1}}$. The spectra of the samples
mentioned below were measured under similar conditions with the detector
operated at integration times of 15 min.

\subsection{Spectra normalization}

To get a good signal to noise and avoid risk of crystallization calls for
compromises. The laser spot used had a diameter of 2$\mu$m (Renishaw setup)
and 2.7$\mu$m (BGU setup) and the question of sample uniformity being
typically 1x1cm$^{\text{2}}$, had to be tested. Taking advantage of the
computerized Invidia (Renishaw) features, several runs of 25 spots on a sample
were taken for statistics purposes. These 25 scans were taken consecutively on
a 0.1x0.1mm square (5 rows, 5 columns). The Raman signal in the interval
440-450cm$^{\text{-1}}$ was averaged for each of the 25 traces and the
histogram of these values is shown in Fig.12:%

%TCIMACRO{\FRAME{ftbpFU}{3.3122in}{4.4278in}{0pt}{\Qcb{{}Top: Histogram of the
%25 local-intensities of IR signal taken on a 55nm In$_{\text{x}}$O film (see
%text). Bottom: Raman spectra taken from two local readings with extreme
%low-high intensities in the histogram and illustrating that they only differ
%by a numerical factor.}}{}{fig_sm4.eps}{\special{ language "Scientific Word";
%type "GRAPHIC";  maintain-aspect-ratio TRUE;  display "USEDEF";
%valid_file "F";  width 3.3122in;  height 4.4278in;  depth 0pt;
%original-width 7.1581in;  original-height 9.5683in;  cropleft "0";
%filename 'Fig_SM4.eps';file-properties "XNPEU";}} }%
%BeginExpansion
\begin{figure}[ptb]%
\centering
\includegraphics[
height=4.4278in,
width=3.3122in
]%
{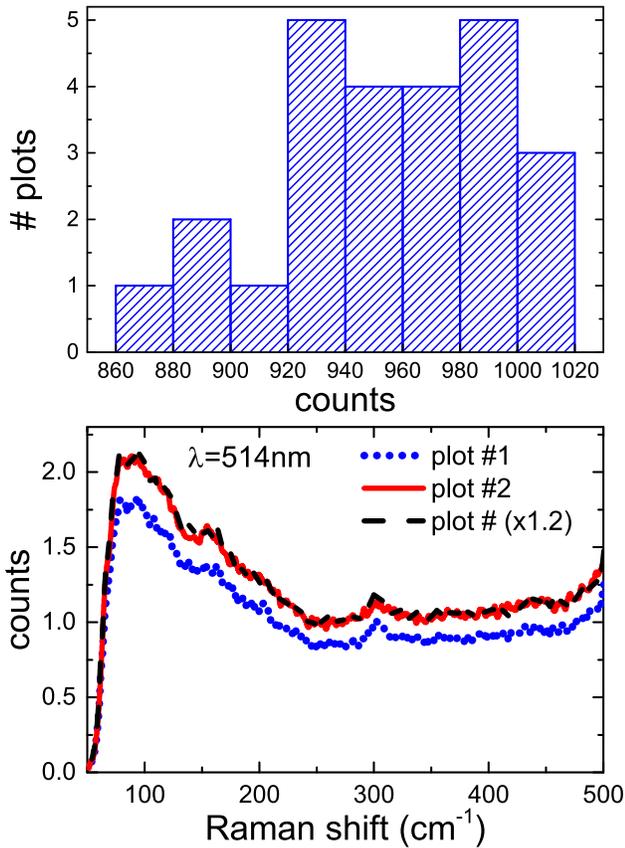}%
\caption{{}Top: Histogram of the 25 local-intensities of IR signal taken on a
55nm In$_{\text{x}}$O film (see text). Bottom: Raman spectra taken from two
local readings with extreme low-high intensities in the histogram and
illustrating that they only differ by a numerical factor.}%
\end{figure}
%EndExpansion

The figure also includes the spectra for two traces for which the variation in
the magnitude of the signal is the greatest and it is illustrated that they
can be made to overlap by multiplying by a constant factor. On basis of such
measurements we opted for normalizing the curves using I$_{\text{0}}$ as the
reference (see Fig.2 of the main text).

\end{document}